\pdfoutput=1

\documentclass[11pt,twoside,a4paper,cmspaper,final,collab]{cms-tdr}
\def\svnVersion{407996}\def\cmsCernNoTag{CERN-EP/2016-237}\def\cmsCernDate{\today}\def\cmsMessage{Published in the Journal of High Energy Physics as \href{http://dx.doi.org/10.1007/JHEP05(2017)013}{\doi{10.1007/JHEP05(2017)013.}}}

\begin{document}\cmsNoteHeader{BPH-14-008}

\hyphenation{had-ron-i-za-tion}
\hyphenation{cal-or-i-me-ter}
\hyphenation{de-vices}
\RCS$Revision: 407911 $
\RCS$HeadURL: svn+ssh://svn.cern.ch/reps/tdr2/papers/BPH-14-008/trunk/BPH-14-008.tex $
\RCS$Id: BPH-14-008.tex 407911 2017-06-01 14:33:16Z mhaytmyr $
\newlength\cmsFigWidth
\ifthenelse{\boolean{cms@external}}{\setlength\cmsFigWidth{0.85\columnwidth}}{\setlength\cmsFigWidth{0.4\textwidth}}
\ifthenelse{\boolean{cms@external}}{\providecommand{\cmsLeft}{top}}{\providecommand{\cmsLeft}{left}}
\ifthenelse{\boolean{cms@external}}{\providecommand{\cmsRight}{bottom}}{\providecommand{\cmsRight}{right}}
\cmsNoteHeader{BPH-14-008}
\title{Observation of $\Upsilon(\mathrm{1S})$ pair production in proton-proton collisions at $\sqrt{s} = 8\TeV$}

\date{\today}
\abstract{
Pair production of $\PgUa$ mesons is observed at the LHC in proton-proton collisions at $\sqrt{s} = 8\TeV$ by the CMS experiment in a data sample corresponding to an integrated luminosity of 20.7\fbinv. Both $\PgUa$ candidates are fully reconstructed via their decays to $\mu^{+}\mu^{-}$. The fiducial acceptance region is defined by an absolute $\PgUa$ rapidity smaller than 2.0. The fiducial cross section for the production of $\PgUa$ pairs, assuming that both mesons decay isotropically, is measured to be $68.8 \pm 12.7\stat \pm 7.4\syst \pm 2.8 \, (\mathcal{B})\unit{pb}$, where the third uncertainty comes from the uncertainty in the branching fraction of $\PgUa$ decays to $\mu^{+}\mu^{-}$. Assuming instead that the $\PgUa$ mesons are produced with different polarizations leads to variations in the measured cross section in the range from $-38\%$ to $+36\%$.
}
\hypersetup{%
pdfauthor={CMS Collaboration},%
pdftitle={Observation of Y(1S) pair production in pp collisions at sqrt(s) = 8 TeV},%
pdfsubject={CMS},%
pdfkeywords={CMS, physics, quarkonium production}}
\maketitle
\clearpage
\clearpage

\clearpage
\newpage
\section{Introduction}\label{sec:introduction}
\indent
The measurement of quarkonium pair production in proton-proton (pp) collisions 
provides insight into the underlying mechanism of particle production. 
The production mechanisms of heavy-quarkonium pairs such as \JPsi and $\PgU$ are especially valuable as they probe 
these interactions in both perturbative and nonperturbative regimes~\cite{Andronic:2015wma}. In this paper, the first cross 
section measurement of $\PgUa$ pair production, reconstructed in the four-muon final state, in pp collisions from the 
CERN LHC at $\sqrt{s}= 8\TeV$ is reported. 

Proton collisions can be described by parton models~\cite{Ioffe:1985ep} in quantum chromodynamics (QCD). 
In this framework, each colliding hadron is characterized as a collection of free 
elementary constituents. Because of the composite nature of hadrons, in a single hadron-hadron collision two 
partons often undergo a single interaction (single-parton scattering, SPS). However, it is also possible that 
multiple distinct interactions (multiple-parton interactions, MPIs) occur, the simplest 
case being double-parton scattering (DPS). The SPS mechanism for heavy-quarkonium pair production 
can be described by a nonrelativistic effective theory~\cite{Bodwin:1995gT} of QCD. However, 
contributions from the DPS mechanism are not addressed in a simple way within the framework of perturbative 
QCD~\cite{Ko:2010xy}, and DPS or MPIs are widely invoked to account for the observations 
that cannot be explained otherwise, such as the rates for multiple heavy-flavor production~\cite{Berger:2009cm}. 
There are still large uncertainties owing to possible higher-order SPS contributions and the limited knowledge of the proton 
transverse profile~\cite{Diehl:2011yj}. Heavy-quarkonium final states are expected to probe the distribution of gluons in a 
proton since their production is dominated by gluon-gluon (gg) interactions~\cite{Baranov:2012re,Berezhnoy:2012tu}. Cross 
section measurements of quarkonium pair production are essential in understanding SPS and DPS contributions and the parton 
structure of the proton. 

The first quarkonium pair production measurement dates back to 1982 when the NA3 experiment at CERN observed direct
production of two \JPsi mesons~\cite{Badier:1982ae} in $\pi^{-}$-platinum interactions with beam energy of 150\GeV and 280\GeV, 
where the main contribution is quark-antiquark annihilation. Early theoretical calculations of the pair production cross section assumed 
only color-singlet states~\cite{Ellis:1976fj,Carlson:1976cd,Kuhn:1979kb}.
However, because of the high parton flux and high center-of-mass energy at the LHC, 
DPS is expected to play a significant role in quarkonium pair production~\cite{Baranov:2011ch}. 

Quarkonium pair production in pp collisions via DPS is assumed to result from two independent SPS 
occurrences~\cite{Baranov:2012re,Baranov:2011ch}. Several DPS production processes, including final states with associated 
jets, are commonly described by an effective cross section ($\sigma_\mathrm{eff}$) that characterizes the transverse area of 
the hard partonic interactions~\cite{Calucci:1999yz,Calucci:1997uw}. It is estimated as the product of
single quarkonium production cross sections divided by the corresponding DPS quarkonium pair cross section~\cite{Baranov:2016mix}. 
Assuming the parton distribution functions are not correlated, $\sigma_\mathrm{eff}$ is expected to be independent of the final 
state and the center-of-mass energy. Recent measurements of $\sigma_\mathrm{eff}$ in final states 
with jets are in the range 12$-$20 mb~\cite{Chatrchyan:2013xxa, Aad:2013bjm, Abazov:2015nnn, Abazov:2014fha}. 
However, measurements of \JPsi pair~\cite{Abazov:2014qba} and \JPsi$+\Upsilon(1\mathrm{S})$~\cite{Abazov:2015fbl} 
production in \Pp\Pap collisions yield values of 4.8 and 2.2 mb, respectively. From a fit to differential cross section 
measurements of \JPsi pair production with CMS~\cite{Khachatryan:2014iia}, Lansberg and Shao~\cite{Lansberg:2014swa} estimate 
an effective cross section of 8.2 mb. In this paper, we report the first observation of $\PgUa$ pair production and 
estimate $\sigma_\mathrm{eff}$ using this result, along with theoretical predictions.  

\section{The CMS detector and muon reconstruction}\label{sec:detector}
\indent
The central feature of the CMS apparatus is a 13\unit{m} long superconducting solenoid of 6\unit{m} 
internal diameter, providing a magnetic field of 3.8\unit{T}. Within the solenoid volume are a silicon pixel 
and strip tracker, a lead tungstate crystal electromagnetic calorimeter, and a brass and scintillator hadron calorimeter,
each composed of a barrel and two endcap sections. 
Extensive forward calorimetry complements the coverage provided by the barrel and endcap detectors. 
Muons are measured in gas-ionization detectors embedded in the steel flux-return yoke outside the solenoid. 
The main subdetectors used for the present analysis are the muon detection system and the silicon tracker.

Charged particle trajectories are reconstructed by the silicon tracker in the pseudorapidity range $\abs{\eta}< 2.5$. 
The innermost component of the tracker consists of three cylindrical layers of pixel detectors in the barrel region and 
two endcap disks at each end of the barrel. The strip tracker has 10 barrel layers and 12 endcap disks at each end of the 
barrel. There are 66 million $100\times150\mum^2$ silicon pixels and more than 9 million silicon strips. 
Strip pitches vary between 80\mum and 120\mum in the inner barrel, while the inner disk sensors have a pitch between 
100\micron and 141\micron. For charged particles of transverse momentum $1 < \pt< 10$\GeV and $\abs{\eta} < 1.4$, 
the relative track resolution is typically 1.5\% in \pt and 25$-$90 (45$-$50)\mum in the transverse (longitudinal) impact 
parameter~\cite{Chatrchyan:2014fea}.

The outermost part of the CMS detector is the muon system, which covers $|\eta|<$ 2.4. It consists 
of four layers of detection planes made of drift tubes, cathode strip chambers, and resistive-plate chambers. 
Muon reconstruction and identification begins within the muon system using hits in the muon chambers. 
A matching algorithm makes the best association between a track segment in the muon system and a track segment in the inner tracker. 
A track fit is performed combining all hits from both subsystems to select high-purity muon candidates.

Data are collected with a two-level trigger system. The first level of the trigger system, composed of custom hardware 
processors, uses information from the calorimeters and muon detectors to select events that pass the trigger requirements in a fixed time 
interval of about 4\mus, of which 1\mus is available for data processing.
The second stage, the high-level trigger (HLT), is a processor farm that further reduces the 
event rate from around 100\unit{kHz} to less than 1\unit{kHz} before data storage. 
The HLT has full access to all the event information, including tracking, and
therefore selections based on algorithms similar to those applied offline are used.

A more detailed description of the CMS detector, together with a definition of the coordinate system used and the relevant 
kinematic variables, can be found in ref.~\cite{Chatrchyan:2008aa}.

\section{Data, simulation, and event selection}\label{sec:eventselection}
\indent
The data sample used in this analysis was collected with the CMS detector at the LHC in proton-proton
collisions at a center-of-mass energy of 8 TeV and corresponds to an integrated luminosity of 20.7 fb$^{-1}$. 
The average number of simultaneous pp collisions in a bunch crossing (pileup) during this run was 16.  

Two separate simulated samples that account for the different production mechanisms of
$\PgUa$ meson pairs are used to validate the efficiency and acceptance correction methods and parameterize the kinematic 
distributions of the data. Strongly correlated $\PgUa$ meson pairs produced via SPS are generated with the
{\sc cascade 2.3.13} simulation program~\cite{Jung:2010si}, while less-correlated $\PgUa$ meson pairs produced via DPS are generated with 
the {\sc pythia 8.1.58} package~\cite{Sjostrand:2007gs}. The {\sc cascade} event simulation uses an off-shell matrix-element description of the 
g*g* $\rightarrow \PgUa\PgUa$ production and includes the \kt-dependent parton distribution function JH-2013-set2~\cite{Hautmann:2013tba}. 
The parton evolution follows the CCFM prescription~\cite{Ciafaloni:1987ur,Catani:1989sg,Marchesini:1994wr}, which forms a bridge between the 
DGLAP and BFKL parton resummation models. The CMS detector response is simulated with the \GEANTfour toolkit~\cite{Agostinelli:2002hh}. 
These samples are simulated with the appropriate conditions for the analysed data, including the effects of alignment, efficiency, and pileup.
 
For the rest of the paper, the notation $\PgU\PgU$ will be used to denote any pair-wise combination of $\PgUa$, $\PgUb$, and $\PgUc$ mesons. 
Candidate $\PgU\PgU$ events are required to have four or more muons that pass the first-level muon trigger, of which at
least three must be identified as muons by the HLT. There must be at least one pair of oppositely charged muons that have an invariant mass
$M_{\mu\mu}$ between 8.5 and 11\GeV, and a vertex fit chi-squared ($\chi^2$) probability greater than 0.5\%, 
as determined from a Kalman-filter algorithm~\cite{Fruhwirth:1987fm}.

The offline selection of $\Upsilon\Upsilon$ candidates begins with a search for four muons with the sum
of their charges equal to 0. Selected muons are further required to pass the following quality criteria: 
tracks of muon candidates must have at least six hits in the silicon tracker, at least one hit in the pixel detector, and they must 
match at least one muon segment in the muon detector. Loose selection cuts are applied on the 
longitudinal and transverse impact parameters of the muons to reject muons from cosmic rays and weakly decaying hadrons.
To ensure a nearly uniform single-muon acceptance and a well-defined kinematic region, the muon candidates must have $\pt^{\mu}>3.5\GeV$ 
and $|\eta^{\mu}|<2.4$. 

To reconstruct the $\PgUa\PgUa \rightarrow 4\mu$ decay, a kinematic fit~\cite{Prokofiev:2004am} is 
performed on oppositely charged muon pairs and then on the four muons. 
The fit incorporates the decay and production kinematic properties to reconstruct the full decay chain of the user-defined process. 
To be compatible with the trigger requirements, each reconstructed $\PgU$ candidate is required to have a vertex 
fit $\chi^2$ probability larger than 0.5\%, as determined by kinematic fitting, and an invariant mass between 8.5 and 11\GeV. 
To suppress contributions from pileup events, all four muons are also required to be associated with a common vertex 
having a fit $\chi^2$ probability larger than 5\%.

In order to measure the cross section in a well-define kinematic region, both $\PgU$ candidates must have $|y^{\PgU}|<2.0$, where $y$ is the 
rapidity. Some of the events containing more than four muons result in multiple $\Upsilon\Upsilon$ candidates per event since multiple dimuon 
pairs can pass the trigger and selection requirements. These events constitute about 4\% of the selected events and are 
discarded from further analysis. After all selection criteria are applied, a total of 313 $\PgU\PgU$ candidates are found. 
The major contribution of background comes from miscombined muons from 
Drell--Yan production and semileptonic $ \PQb \rightarrow \mu + \mathrm{X}$ or  $\PQc \rightarrow \mu + \mathrm{X}$ decays that pass the trigger and selection requirements.

\begin{figure}[bt]
\begin{center}
\includegraphics[width=0.60\textwidth]{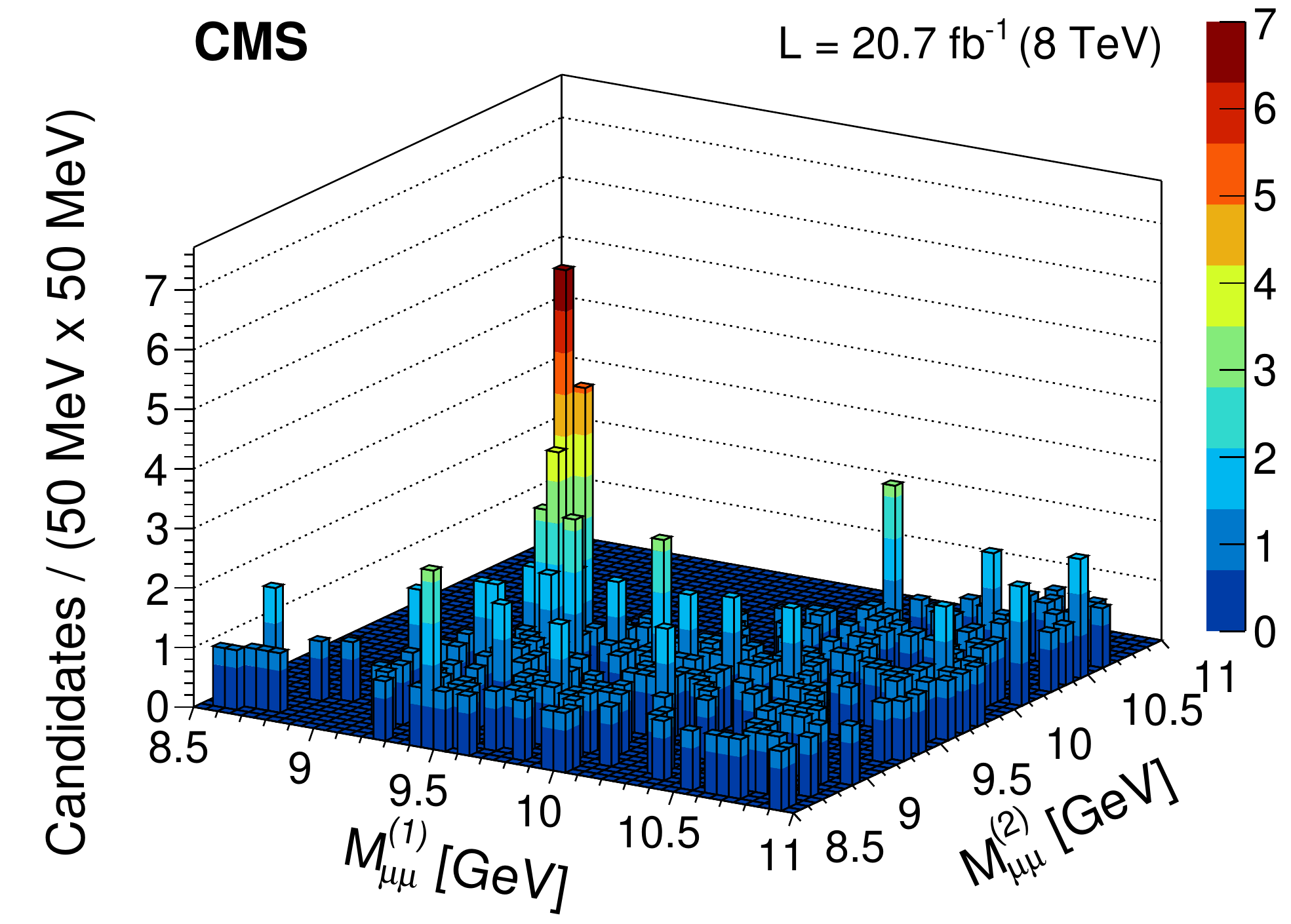}
\caption{Two-dimensional distribution of $M^{(1)}_{\mu\mu}$ vs. $M^{(2)}_{\mu\mu}$.}
\label{scatter_plot}
\end{center}
\end{figure}

To extract the signal yield of $\PgUa\PgUa$ events, a two-dimensional (2D) unbinned extended maximum-likelihood fit to the invariant 
masses of the two reconstructed $\mu^{+}\mu^{-}$ combinations is used. Two kinematic variables are employed to discriminate the 
$\PgUa\PgUa$ signal from the background: the invariant mass of the higher-mass $\Upsilon$ candidate ($M^{(1)}_{\mu\mu}$) and 
the invariant mass of the lower-mass $\Upsilon$ candidate ($M^{(2)}_{\mu\mu}$). 
This choice of variables helps to resolve the ambiguity of having a dimuon pair in which an $\Upsilon$(2S) candidate appears. 
There is no visible signal for $\Upsilon$(3S), which is not considered further in this study. 
Figure~\ref{scatter_plot} shows the distribution of $M^{(1)}_{\mu\mu}$ versus $M^{(2)}_{\mu\mu}$ after all selection criteria are applied.
Five categories of events are considered to represent the sample of events:
\begin{itemize}
\item [1.] events containing a genuine $\PgUa\PgUa$ pair (labeled signal),
\item [2.] events containing a genuine $\PgUb\PgUa$ pair (labeled $\PgUb$--$\PgUa$),
\item [3.] events containing a genuine $\PgUa$ and two unassociated muons that contribute to the $M^{(2)}_{\mu\mu}$ 
distribution (labeled $\PgUa$--combinatorial),
\item [4.] events containing a genuine $\PgUb$ and two unassociated muons that contribute to the $M^{(2)}_{\mu\mu}$ 
distribution (labeled $\PgUb$--combinatorial), and    
\item [5.] events containing four unassociated muons (labeled combinatorial).
\end{itemize}
The contributions from other categories (combinatorial--$\PgUa$ and combinatorial--$\Upsilon$(2S)), 
where two unassociated muons form the higher-mass $\PgU$ candidate, are found to be consistent with zero. 
There is no indication of $\PgUb\PgUb$ events.

Each opposite-sign dimuon invariant mass distribution contains the contributions from $\PgU$ candidates and a nonresonant background.
The $\PgU$ signal is modeled as a sum of two Crystal Ball functions~\cite{Oreglia:1980cs} 
with a common mean, labeled DCB, to capture the 
variation in dimuon resolution as a function of $y^{\PgU}$. The DCB parameters are extracted from the signal SPS simulated sample and fixed 
in the fit to the data. The combinatorial background components are modeled with first-order Chebyshev polynomials (Poly). 
The distributions of the complete combinatorial background are found to be compatible with those of like-sign dimuon combinations.
The yields of the five event categories and the two slope parameters for the combinatorial background are the free parameters in the fit to data. 
The likelihood function $\ell_{k}$ for an event $k$ (ignoring the normalisation term for simplicity) is given as:
\begin{equation}
\begin{split}
\ell_{k}&	= N_\mathrm{signal} \, [\,\mathrm{DCB}_{\Upsilon\mathrm{(1S)}}(M^{(1)}_{\mu\mu}) \, \mathrm{DCB}_{\Upsilon\mathrm{(1S)}}(M^{(2)}_{\mu\mu})\,] \\
&       + N_{\Upsilon\mathrm{(2S)}-\Upsilon\mathrm{(1S)}}\,[\,\mathrm{DCB}_{\Upsilon\mathrm{(2S)}}(M^{(1)}_{\mu\mu}) \, \mathrm{DCB}_{\Upsilon\mathrm{(1S)}}(M^{(2)}_{\mu\mu})\,]\\        
&       + N_{\Upsilon\mathrm{(1S)-combinatorial}}\,[\,\mathrm{DCB}_{\Upsilon\mathrm{(1S)}}(M^{(1)}_{\mu\mu})\, \mathrm{Poly}(M^{(2)}_{\mu\mu})\,]\\
&       + N_{\Upsilon\mathrm{(2S)-combinatorial}}\,[\,\mathrm{DCB}_{\Upsilon\mathrm{(2S)}}(M^{(1)}_{\mu\mu})\, \mathrm{Poly}(M^{(2)}_{\mu\mu})\,]\\          
&       + N_\mathrm{combinatorial}\,[\,\mathrm{Poly}(M^{(1)}_{\mu\mu})\, \mathrm{Poly}(M^{(2)}_{\mu\mu})\,].
\end{split}
\end{equation}
The yields $N_{i}$ for the five categories are determined using the RooFit package~\cite{Verkerke:2003ir} by minimizing the quantity $-$ln $L$, 
where $L = \prod_{k} \ell_{k}$ is the total extended likelihood function. 
The extracted yields with this procedure are $N_{\text{signal}} = 38\pm7$ and $N_{\PgUb-\PgUa} = 13^{+6}_{-5}$. 
\begin{figure}[!tb]
\begin{center}
\includegraphics[width=0.48\textwidth]{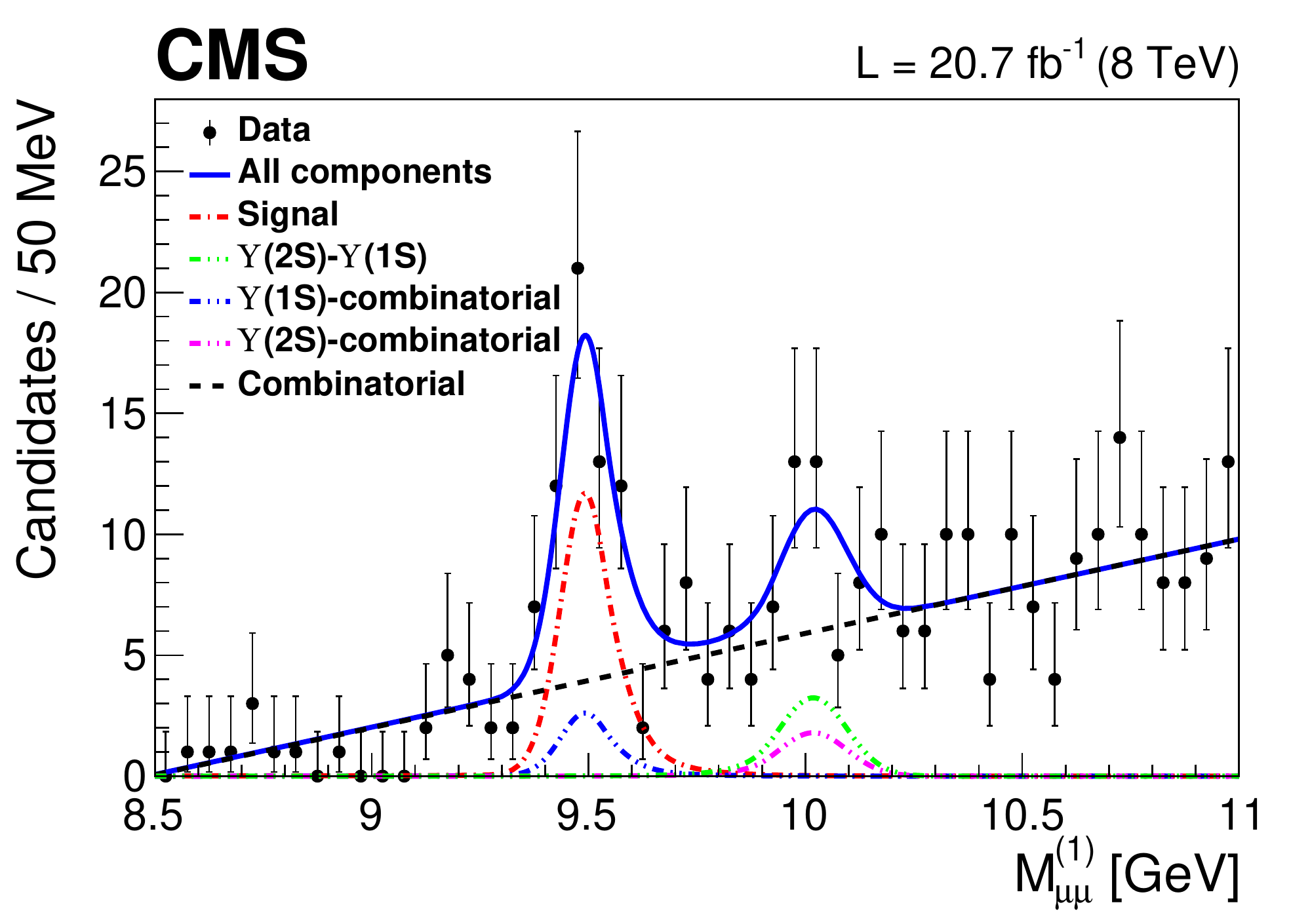}
\includegraphics[width=0.48\textwidth]{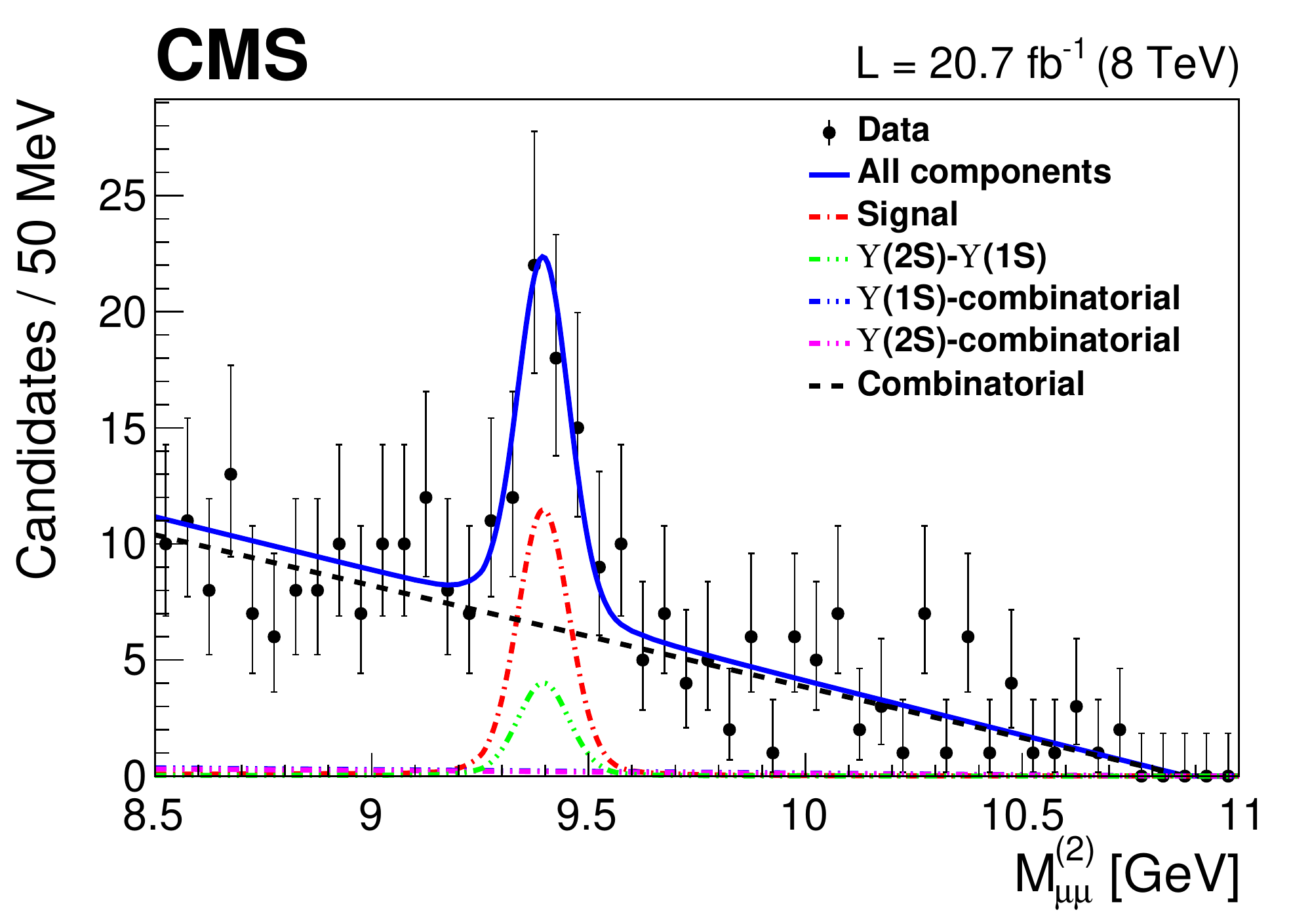}
\caption{Invariant mass distributions of the higher-mass muon pair (left)
and the lower-mass muon pair (right). The data are shown by the points. The different curves show the contributions of the 
various event categories from the fit. }
\label{2d_fit_result}
\end{center}
\end{figure}
To evaluate the significance of the $\PgUa\PgUa$ signal, the consistency of the data is checked relative to
two hypotheses: a background-only hypothesis (the categories containing $\Upsilon$(2S)--$\PgUa$, $\PgUa$--combinatorial, 
$\Upsilon$(2S)--combinatorial, and combinatorial), and a signal hypothesis (the categories containing background and signal). 
The maximum-likelihood values returned from the fits to the signal and background-only hypotheses are denoted by $L_S$ and $L_0$, respectively. 
The resulting statistical significance of the signal is calculated as $\sqrt{\smash[b]{-2 \mathrm(L_S/L_0)}}$~\cite{Wilks:1938dza}, 
and is found to be well above five standard deviations. The statistical significance of the $\PgUb$--$\PgUa$ peak, evaluated using the same 
procedure, is 2.6 standard deviations. The invariant mass distributions of the higher- and lower-mass muon pairs in 
the data, with the result of the 2D fit superimposed, are shown in figure~\ref{2d_fit_result}. 
The fit is validated by applying it to simulated samples generated using the probability distribution 
functions (pdf) that describe the data. No bias in the yields is found, and the likelihood value 
from the fit to the data closely agrees with the mean likelihood value obtained from the simulated samples.

\section{Efficiency and acceptance}\label{sec:effacc}
\indent
The muon acceptance is defined by the geometry of the CMS detector and the kinematic requirements of the analysis, which are 
established based on studies of simulated $\PgUa\PgUa$ events.

The kinematic variable distributions of the $\PgUa\PgUa$ system change with the underlying production mechanism. 
Therefore, to minimize the model dependence, the acceptance and efficiency corrections are calculated on an event-by-event
basis using the measured $\Upsilon$ meson and muon momenta. This method was utilized in a similar CMS 
analysis~\cite{Khachatryan:2014iia}, and is intended to minimize the model dependence of the correction factors. 

The probability of an event passing the muon acceptance requirements is obtained by repeatedly generating
the four-momenta of the four decay muons of the measured $\Upsilon$ pairs in the event. The acceptance correction for each selected event 
is calculated separately by simulating a large sample of such decays. The acceptance correction factor, $a_{i}$ for an event $i$, is defined 
as the number of times all four muons pass the acceptance criteria divided by the total number of trials. The angular distribution of the decay 
muons with respect to the direction of flight of the parent $\PgUa$ meson, in the $\PgUa$ meson rest frame, is assumed to be isotropic. 
Different polarization scenarios of $\PgUa$ mesons are considered and discussed later. 

The efficiency correction for the four-muon system is determined with a data-embedding method that repeatedly substitutes the measured muon 
four-momenta into different simulated events, which are then subjected to the complete CMS 
detector simulation and reconstruction chain. Each event is simulated with the same pileup scenario as in the corresponding data event. 
The efficiency correction factor $\epsilon_{i}$ for the $i^{th}$ event in the data is the number of simulated events that pass the offline 
reconstruction and trigger requirements, divided by the total number of simulated events. For each data event, 1000 different 
simulated events are generated. The number of efficiency-corrected signal events, $N_{\mathrm{true}}$, is then given as 
$\sum_{i} 1/\epsilon_{i}$, where the summation is over the data events that satisfy all the selection requirements. 

Because of the finite $\mu^{+}\mu^{-}$ mass resolution, typically near the kinematic acceptance of the detector, the measured momenta of 
reconstructed $\PgU$ candidates will be distorted from the original sample. To account for this, scaling
factors are determined for both the SPS and DPS simulated samples. First, an average efficiency using a data-embedding method, 
$\overline{\epsilon}_{1}$, is calculated as the number of events surviving 
the trigger and reconstruction requirements, divided by the number of efficiency-corrected events $N_{\mathrm{true}}$. 
Alternatively, an average efficiency using the generator information, $\overline{\epsilon}_{2}$, is determined as the number of events 
satisfying the trigger and reconstruction criteria, divided by the number of events generated within the muon and $\PgU$ meson acceptance 
of the detector. Compared to the first method, the latter efficiency is based on the originally generated muon momenta. 
Then, the scaling factor is defined as the ratio $\overline{\epsilon}_{2}$ / $\overline{\epsilon}_{1}$. 
Scaling factors of 98\% and 96\% are determined for the SPS and DPS models, respectively.
The average of these scaling factors is applied to the data-embedding efficiency. 

\section{Systematic uncertainties}\label{sec:systematic}
\indent
Several sources of systematic uncertainty in the cross section measurement are considered and discussed below. 

To estimate the uncertainty caused by the imperfect modeling of the data, several variations are checked: (i) the parameters for the $\PgUa$ 
resonance shape that are fixed for the maximum-likelihood fit are varied by their uncertainty, as determined from the simulated samples;
(ii) the $\PgUa$ resonance shape is alternatively parameterized with a relativistic Breit--Wigner convolved with a Gaussian function; 
(iii) $\PgUa$ mesons are alternatively parameterized using signal parameters obtained from the DPS simulation samples. 
The largest difference in the signal yield among these fits of 7.9\% is taken as the corresponding systematic uncertainty. 

Evaluation of the trigger and reconstruction efficiencies for $\PgU$ pair production rely on detector simulation. To estimate the systematic 
uncertainty owing to the detector simulation, the four-muon efficiency is also calculated from the single-muon efficiencies measured in data 
with a "tag-and-probe" method~\cite{Chatrchyan:2012xi} that utilizes data samples containing a single \JPsi meson decaying into a muon pair.
To calculate the four-muon efficiency using the tag-and-probe method, the correlations among the muons is separately derived from the simulation.
A total correlation factor, determined from simulation as a function of the \pt and rapidity of the four-muon system, is multiplied by the 
product of the four single-muon efficiencies, also measured versus $\eta$ and \pt. The difference of the corrected signal yield with the 
tag-and-probe method and the data-embedding method of 4.9\% is taken as the systematic uncertainty from the simulation.	

The efficiency correction method is evaluated using the full detector simulation and reconstruction event samples generated according to the 
two production mechanisms: SPS and DPS. The systematic uncertainty from using the efficiency-correction method, a data-embedding efficiency 
correction method is applied to obtain the per-event efficiency, $\epsilon_{i}$. 
The efficiency-corrected yield is compared to the number of generated 
events, and the fractional difference between the two is used to estimate the uncertainty in the correction method. This procedure is repeated 
for both simulated samples separately. The systematic uncertainty is then taken to be the larger of the differences in the two samples and 
is found to be 3.7\%.

Signal SPS and DPS simulated samples are used to estimate the uncertainty in the event-by-event acceptance correction method. 
A sample of $N_{\mathrm{gen}}$ simulated events are subjected to the acceptance criteria. The event-based acceptance correction is then 
applied to the events passing the full selection to arrive at a corrected yield, $N'_{\mathrm{gen}}$. The uncertainty is taken as the relative 
deviation of $N'_{\mathrm{gen}}$ from $N_{\mathrm{gen}}$. The larger deviation of 2.8\% between the SPS and DPS samples is taken as a systematic uncertainty. 

The luminosity delivered to CMS is measured based on cluster counting from the pixel detector. The overall systematic uncertainty in the 
integrated luminosity is estimated to be 2.6\%~\cite{CMS-PAS-LUM-13-001}.

Table~\ref{total_sys_uncertainty} provides a summary of the individual systematic uncertainties. The total systematic uncertainty 
in the $\PgUa\PgUa$ cross section measurement of 10.7\% is calculated as the sum in quadrature of the individual components.
\begin{table}[!htb]
\centering
\topcaption{Summary of the relative systematic uncertainties in the $\PgUa$ pair production cross section.}
\begin{tabular}{c|c}
\hline
Component &  Uncertainty (\%)  \\
\hline
Resonance shape & 7.9 \\
Simulation & 4.9 \\
Efficiency & 3.7 \\
Acceptance & 2.8 \\
Integrated luminosity & 2.6  \\
\hline
Total &  10.7 \\
\hline
\end{tabular}
\label{total_sys_uncertainty}
\end{table}
In this analysis, $\PgUa$ mesons are assumed to decay isotropically. To check the sensitivity of the fiducial cross section measurement 
on the $\PgUa$ meson decay angular distribution, studies are performed for several extreme polarization scenarios. 
With $\theta^{*}$ defined as the angle between the $\mu^{+}$ direction in the $\PgUa$ meson rest frame and the direction 
of the $\PgUa$ in the lab frame, the angular distribution is parameterized as $1+\lambda_{\theta}\cos^{2}(\theta^{*})$, where 
$\lambda_{\theta}$ is the polar anisotropic parameter~\cite{Chatrchyan:2012woa}. The extreme cases $\lambda_{\theta} = 0$ or 
$\lambda_{\theta} = +1$ ($-$1) corresponds to an isotropic $\PgUa$ meson decay or 100\% longitudinal (transverse) polarization.
The polar anisotropic parameter for each $\PgUa$ candidate, labeled as $\lambda_{\theta 1}$ and $\lambda_{\theta 2}$, are varied 
simultaneously, and Table~\ref{polarization_systematics} shows a summary of this study. Compared to an isotropic $\PgUa$ meson decay, 
the corrected yield is 38\% lower for $\lambda_{\theta1} = \lambda_{\theta2} = -1$ and 36\% higher for 
$\lambda_{\theta1} = \lambda_{\theta2} = +1$. 
This variation is not included in the total systematic uncertainty, but is quoted separately.
\begin{table}[!ht]
\centering
\topcaption{Relative change in percent of the acceptance-corrected $\PgUa\PgUa$ yield for different polarization assumptions
with respect to that for $\lambda_{\theta1} = \lambda_{\theta2} = 0$.}
\begin{tabular}{c|cccccccccc}
\hline
$\lambda_{\theta 1}$ & $+$1 &$+$1\phantom{.0} &$+$0.5 & $-$0.5 & $+$0.5 &$-$1 & $+$1 &  $-$0.5 &$-$0.5&$-$1 \\
$\lambda_{\theta 2}$ & $+$1 &$+$0.5&$+$0.5 & $+$0.5 &$-$0.5 &$+$1 & $-$1 &  $-$0.5 & $-$1\phantom{.0} &$-$1 \\
\hline
Change (\%) & $+$36 & $+$26& $+$18& $-$2 & $-$3 & $-$9 &$-$9 & $-$19&$-$29&$-$38 \\
\hline
\end{tabular}
\label{polarization_systematics}
\end{table}

\section{Results}\label{sec:xsec}
\indent
The fiducial cross section of $\PgUa$ pair production, $\sigma_\text{fid}$, is measured for events in which both $\PgUa$ mesons 
have $|y^{\PgU}|<2.0$. It is derived from the measured $\PgUa$ pair yield, $N_\mathrm{signal}$, using the expression:
\begin{equation}\label{xsection_form} 
\sigma_{\mathrm{fid}}  = 
\frac{N_\mathrm{signal} \, \overline{\omega}
}{ 
\mathcal{L} \, [\mathcal{B}(\Upsilon\mathrm{(1S)} \rightarrow \mu ^{+} \mu ^{-})]^{2}}, 
\end{equation}
where $\overline{\omega}$ is the average correction factor that accounts for the inefficiencies stemming from reconstruction, identification, and trigger requirements, as well as the fraction of events lost because of the muon acceptance criteria, $ \mathcal{L} $ is the integrated luminosity, and $\mathcal{B}(\PgUa\rightarrow \mu^{+}\mu^{-})$ $=$ (2.48 $\pm$ 0.05)\% is the world-average branching fraction of the $\PgUa$ to $\mu^{+}\mu^{-}$~\cite{Olive:2016xmw}.

The factor $\overline{\omega}$ is the inverse product of per-event efficiencies, $\epsilon_{i}$, and acceptances, $a_{i}$, averaged over 
the events that pass selection cuts. For the calculation of the total cross section, values of $\mathcal{L} = 20.7\fbinv$,
$N_\mathrm{signal} = 38\pm7$, and $\overline{\omega} = 23.06$ are used. 
Under the assumption that both $\PgUa$ mesons decay isotropically, the total cross section for $\PgUa$ pair production 
is measured to be $\sigma_{\mathrm{fid}} = 68.8 \pm 12.7\stat \pm 7.4\syst \pm 2.8 \, (\mathcal{B})\unit{pb}$, 
where the uncertainties are statistical, systematic, and related to the $\PgUa\rightarrow \mu^{+}\mu^{-}$ branching fraction, respectively.

In quarkonium pair production, the measurement of the effective cross section $\sigma_\mathrm{eff}$ depends on the fraction of DPS,
which is usually estimated either as a residual to the SPS prediction or as the result of a fit to the rapidity
or azimuthal angle difference between quarkonia pairs. To estimate $\sigma_\mathrm{eff}$ using the fiducial $\PgUa$ pair
production cross section $\sigma_{\mathrm{fid}}$, the following formula is used~\cite{Baranov:2012re}:
\begin{equation}\label{sigma_eff_form} 
\sigma_\mathrm{eff}  = \frac{[\sigma(\Upsilon)]^2}{2 \,
f_{\mathrm{DPS}} \, \sigma_{\mathrm{fid}} \,
[\mathcal{B}(\PgUa \rightarrow \mu ^{+} \mu ^{-})]^{2} 
},
\end{equation}
where $\sigma(\PgU)$ is the cross section for single-$\PgUa$ production~\cite{Chatrchyan:2013yna}
extrapolated to the fiducial region ($|y^{\PgU}|<2.0$) and $f_{\mathrm{DPS}}$ is the fraction of the DPS contribution.
To estimate the effective cross section, we use $\sigma(\Upsilon) = 7.5\pm0.6\unit{nb}$ and a value of
$f_{\mathrm{DPS}} \approx 10\%$~\cite{Berezhnoy:2012tu}, which gives $\sigma_\mathrm{eff} \approx 6.6\unit{mb}$. 
The cross section for $\PgUa$ pair production, assuming SPS with feed-down from higher $\PgU$ states,
is 48\unit{pb}~\cite{Berezhnoy:2012tu}. This result, combined with our measurement, gives
$f_{\mathrm{DPS}}\simeq 30\%$, and with this estimation, $\sigma_\mathrm{eff}$ is calculated to be ${\approx} 2.2\unit{mb}$.
These two estimates of $\sigma_\mathrm{eff}$ are consistent with the range of values from the
heavy-quarkonium measurements (2$-$8\unit{mb})~\cite{Abazov:2015fbl,Abazov:2014qba}, but are smaller than that from multijet
studies (12$-$20 mb)~\cite{Chatrchyan:2013xxa, Aad:2013bjm, Abazov:2015nnn, Abazov:2014fha}. The quarkonium final states are dominantly 
produced from gluon-gluon interactions, while the jet-related channels that correspond to higher values of $\sigma_\mathrm{eff}$ are produced 
by quark-antiquark and quark-gluon parton interactions. 
The trend in the measured effective cross section might indicate that the average transverse distance 
between gluons in the proton is smaller than between quarks, or between quarks and gluons.

\section{Summary}\label{sec:summary}

The first observation of $\PgUa$ pair production has been performed with the CMS detector in proton-proton collisions at 
$\sqrt{s}= 8\TeV$ from a sample corresponding to an integrated luminosity of 20.7\fbinv. A signal yield of 
$38 \pm 7$ $\PgUa\PgUa$ events is measured with a significance exceeding five standard deviations. 
Using the measured muon kinematic properties in the data, an event-based acceptance and efficiency-correction method
is applied, minimizing the model dependence of the cross section measurement. The fiducial cross section
of $\PgUa$ pair production measured within the single-$\PgU$ acceptance $|y^{\PgU}|<2.0$ is found to be 
$\sigma_{\mathrm{fid}} = 68.8 \pm 12.7\stat \pm 7.4\syst \pm 2.8 \, (\mathcal{B})\unit{pb}$, where the $\PgUa$ decay 
into muons is assumed to be isotropic. Assuming instead that the $\PgUa$ mesons are produced with different polarizations leads to variations 
in the measured cross section in the range from $-38\%$ to $+36\%$. An effective cross section is also estimated using our 
result that is in agreement with the values from heavy-quarkonium measurements, but is smaller than that from multijet studies.

\begin{acknowledgments}

\hyphenation{Bundes-ministerium Forschungs-gemeinschaft Forschungs-zentren Rachada-pisek} We congratulate our colleagues in the CERN accelerator departments for the excellent performance of the LHC and thank the technical and administrative staffs at CERN and at other CMS institutes for their contributions to the success of the CMS effort. In addition, we gratefully acknowledge the computing centers and personnel of the Worldwide LHC Computing Grid for delivering so effectively the computing infrastructure essential to our analyses. Finally, we acknowledge the enduring support for the construction and operation of the LHC and the CMS detector provided by the following funding agencies: the Austrian Federal Ministry of Science, Research and Economy and the Austrian Science Fund; the Belgian Fonds de la Recherche Scientifique, and Fonds voor Wetenschappelijk Onderzoek; the Brazilian Funding Agencies (CNPq, CAPES, FAPERJ, and FAPESP); the Bulgarian Ministry of Education and Science; CERN; the Chinese Academy of Sciences, Ministry of Science and Technology, and National Natural Science Foundation of China; the Colombian Funding Agency (COLCIENCIAS); the Croatian Ministry of Science, Education and Sport, and the Croatian Science Foundation; the Research Promotion Foundation, Cyprus; the Secretariat for Higher Education, Science, Technology and Innovation, Ecuador; the Ministry of Education and Research, Estonian Research Council via IUT23-4 and IUT23-6 and European Regional Development Fund, Estonia; the Academy of Finland, Finnish Ministry of Education and Culture, and Helsinki Institute of Physics; the Institut National de Physique Nucl\'eaire et de Physique des Particules~/~CNRS, and Commissariat \`a l'\'Energie Atomique et aux \'Energies Alternatives~/~CEA, France; the Bundesministerium f\"ur Bildung und Forschung, Deutsche Forschungsgemeinschaft, and Helmholtz-Gemeinschaft Deutscher Forschungszentren, Germany; the General Secretariat for Research and Technology, Greece; the National Scientific Research Foundation, and National Innovation Office, Hungary; the Department of Atomic Energy and the Department of Science and Technology, India; the Institute for Studies in Theoretical Physics and Mathematics, Iran; the Science Foundation, Ireland; the Istituto Nazionale di Fisica Nucleare, Italy; the Ministry of Science, ICT and Future Planning, and National Research Foundation (NRF), Republic of Korea; the Lithuanian Academy of Sciences; the Ministry of Education, and University of Malaya (Malaysia); the Mexican Funding Agencies (BUAP, CINVESTAV, CONACYT, LNS, SEP, and UASLP-FAI); the Ministry of Business, Innovation and Employment, New Zealand; the Pakistan Atomic Energy Commission; the Ministry of Science and Higher Education and the National Science Centre, Poland; the Funda\c{c}\~ao para a Ci\^encia e a Tecnologia, Portugal; JINR, Dubna; the Ministry of Education and Science of the Russian Federation, the Federal Agency of Atomic Energy of the Russian Federation, Russian Academy of Sciences, and the Russian Foundation for Basic Research; the Ministry of Education, Science and Technological Development of Serbia; the Secretar\'{\i}a de Estado de Investigaci\'on, Desarrollo e Innovaci\'on and Programa Consolider-Ingenio 2010, Spain; the Swiss Funding Agencies (ETH Board, ETH Zurich, PSI, SNF, UniZH, Canton Zurich, and SER); the Ministry of Science and Technology, Taipei; the Thailand Center of Excellence in Physics, the Institute for the Promotion of Teaching Science and Technology of Thailand, Special Task Force for Activating Research and the National Science and Technology Development Agency of Thailand; the Scientific and Technical Research Council of Turkey, and Turkish Atomic Energy Authority; the National Academy of Sciences of Ukraine, and State Fund for Fundamental Researches, Ukraine; the Science and Technology Facilities Council, UK; the US Department of Energy, and the US National Science Foundation.

Individuals have received support from the Marie-Curie program and the European Research Council and EPLANET (European Union); the Leventis Foundation; the A. P. Sloan Foundation; the Alexander von Humboldt Foundation; the Belgian Federal Science Policy Office; the Fonds pour la Formation \`a la Recherche dans l'Industrie et dans l'Agriculture (FRIA-Belgium); the Agentschap voor Innovatie door Wetenschap en Technologie (IWT-Belgium); the Ministry of Education, Youth and Sports (MEYS) of the Czech Republic; the Council of Science and Industrial Research, India; the HOMING PLUS program of the Foundation for Polish Science, cofinanced from European Union, Regional Development Fund, the Mobility Plus program of the Ministry of Science and Higher Education, the National Science Center (Poland), contracts Harmonia 2014/14/M/ST2/00428, Opus 2013/11/B/ST2/04202, 2014/13/B/ST2/02543 and 2014/15/B/ST2/03998, Sonata-bis 2012/07/E/ST2/01406; the Thalis and Aristeia programs cofinanced by EU-ESF and the Greek NSRF; the National Priorities Research Program by Qatar National Research Fund; the Programa Clar\'in-COFUND del Principado de Asturias; the Rachadapisek Sompot Fund for Postdoctoral Fellowship, Chulalongkorn University and the Chulalongkorn Academic into Its 2nd Century Project Advancement Project (Thailand); and the Welch Foundation, contract C-1845.
\end{acknowledgments}

\clearpage
\bibliography{auto_generated}
\cleardoublepage \appendix\section{The CMS Collaboration \label{app:collab}}\begin{sloppypar}\hyphenpenalty=5000\widowpenalty=500\clubpenalty=5000\textbf{Yerevan Physics Institute,  Yerevan,  Armenia}\\*[0pt]
V.~Khachatryan, A.M.~Sirunyan, A.~Tumasyan
\vskip\cmsinstskip
\textbf{Institut f\"{u}r Hochenergiephysik,  Wien,  Austria}\\*[0pt]
W.~Adam, E.~Asilar, T.~Bergauer, J.~Brandstetter, E.~Brondolin, M.~Dragicevic, J.~Er\"{o}, M.~Flechl, M.~Friedl, R.~Fr\"{u}hwirth\cmsAuthorMark{1}, V.M.~Ghete, C.~Hartl, N.~H\"{o}rmann, J.~Hrubec, M.~Jeitler\cmsAuthorMark{1}, A.~K\"{o}nig, I.~Kr\"{a}tschmer, D.~Liko, T.~Matsushita, I.~Mikulec, D.~Rabady, N.~Rad, B.~Rahbaran, H.~Rohringer, J.~Schieck\cmsAuthorMark{1}, J.~Strauss, W.~Waltenberger, C.-E.~Wulz\cmsAuthorMark{1}
% ...
\vskip\cmsinstskip
% ...
1:~~Also at Vienna University of Technology, Vienna, Austria\\
% ...

\end{sloppypar}
\end{document}